\documentclass[convention]{aesconf2}

\usepackage[utf8]{inputenc}

\usepackage{microtype}

\usepackage[numbers,square]{natbib}

\usepackage{amsmath,amssymb,amsfonts,bm}
\usepackage[ruled, lined, linesnumbered, commentsnumbered, longend]{algorithm2e}
\usepackage{algpseudocode}

\usepackage{yhmath}

\usepackage{booktabs}
\usepackage{color}
\usepackage{url}
\usepackage[bottom=1.38in, left=0.71in, right=0.71in]{geometry}

\newcommand{\VEC}[1]{\boldsymbol{#1}}
\newcommand{\MAT}[1]{\boldsymbol{#1}}
\newcommand{\FIELD}[1]{\mathbb{#1}}
\newcommand{\PR}[1]{\left ( {#1} \right ) }
\newcommand{\BK}[1]{\left [ {#1} \right ] }
\newcommand{\CR}[1]{\left \{ {#1} \right \} }

\newcommand{\MOD}[2]{ {#1} \textrm{ mod } {#2} }
\newcommand{\ABS}[1]{\left | {#1} \right |}

\newcommand{\EXPECTATION}[2]{\mathbb{E}_{#1} \BK{#2}}
\newcommand{\NORM}[1]{\left\lVert#1\right\rVert}

\newcommand{\ARGMAX}[1]{ \arg\max_{#1}}

\DeclareMathSymbol{\minus}{\mathbin}{AMSa}{"39}

\DeclareMathOperator{\atantwo}{atan2}

\DeclareMathOperator*{\argmin}{arg\,min}

\DeclareFontFamily{OMX}{yhex}{}
\DeclareFontShape{OMX}{yhex}{m}{n}{<->yhcmex10}{}
\DeclareSymbolFont{yhlargesymbols}{OMX}{yhex}{m}{n}
\DeclareMathAccent{\ARC}{\mathord}{yhlargesymbols}{"F3}

\SetKwInput{KwRequire}{Require}

\title{Optimized Loudspeaker Panning for Adaptive Sound-Field Correction and Non-stationary Listening Areas}

\author{Yuancheng Luo}

\affil{Audio Technology, Amazon Inc.}

\expresspapernumber{65}

\correspondence{Yuancheng Luo}{luoyuancheng@gmail.com}

\lastnames{Luo}

\shorttitle{Optimized Loudspeaker Panning}

\begin{document}

\twocolumn[
\maketitle 

\begin{onecolabstract}
Surround sound systems commonly distribute loudspeakers along standardized layouts for multichannel audio reproduction. However in less controlled environments, practical layouts vary in loudspeaker quantity, placement, and listening locations / areas. Deviations from standard layouts introduce sound-field errors that degrade acoustic timbre, imaging, and clarity of audio content reproduction. This work introduces both Bayesian loudspeaker normalization and content panning optimization methods for sound-field correction. Conjugate prior distributions over loudspeaker-listener directions update estimated layouts for non-stationary listening locations; digital filters adapt loudspeaker acoustic responses to a common reference target at the estimated listening area without acoustic measurements. Frequency-domain panning coefficients are then optimized via sensitivity / efficiency objectives subject to spatial, electrical, and acoustic domain constraints; normalized and panned loudspeakers form virtual loudspeakers in standardized layouts for accurate multichannel reproduction. Experiments investigate robustness of Bayesian adaptation, and panning optimizations in practical applications.
\end{onecolabstract}
]

\setlength{\belowdisplayskip}{6pt}
\setlength{\belowdisplayshortskip}{6pt}
\setlength{\abovedisplayskip}{6pt}
\setlength{\abovedisplayshortskip}{6pt}

\setlength{\belowcaptionskip}{-2.0pt}


\section{Introduction}
\label{sec:intro}

Surround sound systems for multichannel audio reproduction have risen in popularity in home theater setups that accommodate proper loudspeaker selection, layout, acoustic room treatment, and calibration established by the international telecommunication union (ITU) standards \citep{ITU_755_4}. Conversely, the same accommodations present a barrier to entry for extemporary arrangements where loudspeakers differ in quality and placement, and operate in changeable listening locations / areas, and reverberant environments. Deviating from the standards degrade accurate reproduction of multichannel audio content as intended by the content authors. Therefore, methods from sound-field control and reconstruction correct for the effects of irregular loudspeaker placements and room reverberation in the listening area via acoustic measurement system inversion \citep{de2013analysis, lopez2005room, brannmark2013compensation}, and modal / planewave decomposition \citep{jin2015theory, talagala2014efficient, kolundzija2011reproducing}; such methods however are inapplicable when acoustic measurements remain unavailable. 

\enlargethispage{4\baselineskip}

In the absence of acoustic measurements, other sensing modalities can infer the loudspeaker layout and listening area location. Inertial measurement unit \citep{poulose2019indoor, won2009kalman} and bluetooth low energy \citep{danics2017model, wang2013bluetooth} indoor tracking can estimate changes in loudspeaker position and orientation. Ultrasound \citep{kim2016accurate}, camera, and video can track in-room listener and loudspeaker positions within fields-of-view. Such meta-data yields a 2D layout of the estimated loudspeaker placements, listening location, and a front direction. We therefore reproduce multichannel content at the listener's area by incorporating Bayesian uncertainty of the estimated layout inputs with loudspeaker distance and orientation  normalization \citep{greenfield1991efficient, makivirta2018modeling} to the listener, and then reformulate conventional amplitude panning methods \citep{pulkki1997virtual, pulkki2001spatial, pulkki1999uniform} in terms of constrained optimization along joint spatial \cite{franck2017optimization, LUO_2023_UPMIX}, electrical \citep{sadek2004novel}, and acoustical \cite{batke2010investigation} domains. The paper is organized as follows:

\enlargethispage{4\baselineskip}

\textbf{Section \ref{SEC:PT}} introduces our normalization method for aligning loudspeaker acoustic transfer functions in an arbitrary layout to a common axial-reference target at the listener location; acoustic delay and attenuation compensate for varying loudspeaker-listener distances whereas minimum-phase and all-pass factorizations \citep{oppenheim1999discrete} normalize for loudspeaker orientations relative to listener locations. We integrate estimates of the loudspeaker-listener normalization directions via Bayesian posterior updates of a novel circular distribution conjugate prior, and provide a sample calibration for a sequence of normalization angles.
\textbf{Section \ref{SEC:DC}} presents our novel normalized loudspeaker panning optimization, which solves for frequency-dependent magnitude-gains that satisfy spatial vector-bases, electrical headroom, and acoustic power constraints; we augment the former vector-base amplitude panning with slack (VBAPS) to accommodate constraints in electric and acoustic domains. Next, we derive a panning sensitivity / efficiency objective from the augmented form that measures panned-source discreteness, and give equivalent primary and null-space formulations in fewer variables. Planewave acoustic covariances model anechoic to diffuse-field assumptions for variable sized listening areas. Optimal solutions are found via second-order cone program \citep{ALIZADEH_2003_SECOND_CONE}.
\textbf{Section \ref{SEC:EXP}} applies our model to several practical applications of loudspeaker correction under varying constraints. For high loudness targets, we find optimal gains across loudspeakers for overdriven content that maximize source discreteness. For anechoic to diffuse-field environments, we show that our panning optimization solutions converge from discrete panning to Rayleigh quotient maximizers \citep{HORN_1990_MATRIX}. For circular-panning over varying loudspeaker layouts, we evaluate panning sensitivity across azimuth steering-angles and recommend preferred layouts for different number of loudspeakers. \textbf{Section \ref{SEC:DISC}} discusses results and future work.

\section{Loudspeaker Normalization}
\label{SEC:PT}

Let $S(\nu, \theta)$ be the loudspeaker's electrical-acoustical transfer function at frequency $\nu$ measured at $1$ meter distance along azimuth $\theta$ (radians) in the horizontal plane, with the acoustic path-delay removed. Under far-field assumptions, the loudspeaker frequency response attenuates by the inverse-distance and undergoes pure-delay. It is useful to express the far-field transfer function along a listener-centric coordinate frame, which centers the origin at the listener's location and aligns the $+x$ axis with the listener's facing direction. The acoustic transfer function $H_n(\kappa, \VEC{r}) $ at coordinate $\VEC{r} \in \FIELD{R}^{2 \times 1}$ for the $n^{th}$ loudspeaker located at coordinate $\VEC{u}_n  \in \FIELD{R}^{2 \times 1} $ with the orientation unit-vector $\VEC{o}_n  \in \FIELD{R}^{2 \times 1}$ follows
\begin{equation}
\begin{split}
H_n(\nu, \VEC{r})  & = S\PR{\nu, \theta_n (\VEC{r}) }  \frac{ e^{\minus j \kappa \NORM{\VEC{s}_n (\VEC{r}) }  } }{ \NORM{\VEC{s}_n (\VEC{r}) }}, \quad
\kappa  = \frac{2 \pi \nu }{c}, \\
\theta_n(\VEC{r}) & = \cos^{-1}\PR{\frac{\VEC{o}_n^T \VEC{s}_n (\VEC{r}) }{\NORM{\VEC{s}_n (\VEC{r})} } },  \quad 
\VEC{s}_n (\VEC{r})  = \VEC{r} - \VEC{u}_n, 
\end{split}
\label{EQ:PT:ANECHOIC_TF}
\end{equation}
where $\kappa$ is the angular wavenumber, $c$ is the speed of sound in meters/second, $\VEC{s}_n(\VEC{r})$ is the evaluation direction relative to the loudspeaker's location, and $\theta_n(\VEC{r})$ is the evaluation angle relative to the loudspeaker's orientation. We can normalize the loudspeaker's transfer function to approximate the original loudspeaker's response $S(\nu, \theta)$ within a listening window at the listener's location $\VEC{r}=\VEC{0}$.  

\begin{figure}[h]
\begin{center}
  {\includegraphics[width=0.8\columnwidth]{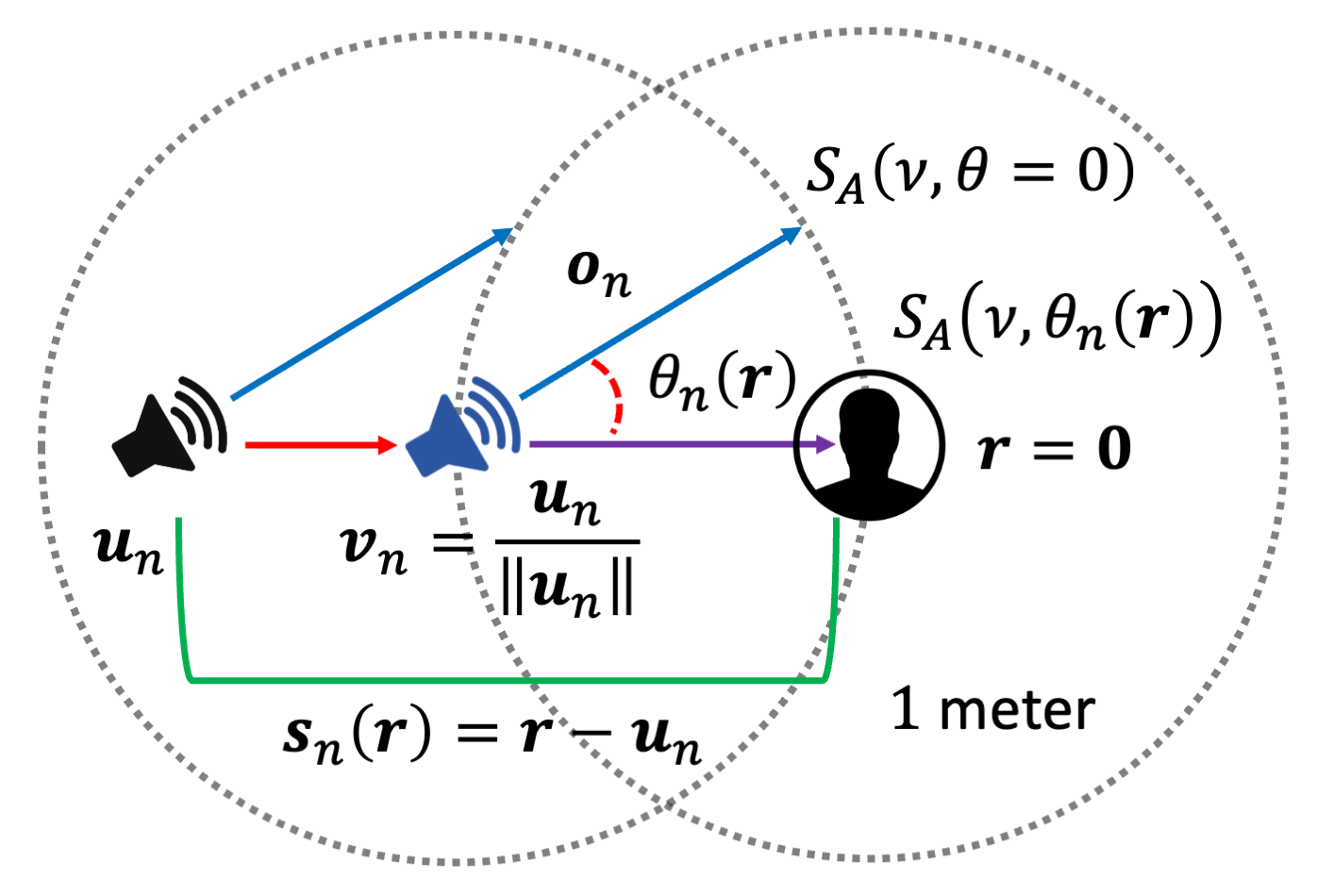}}
\caption{Acoustic transfer function $G_n(\nu)$ in \eqref{EQ:PT:ANECHOIC_TF_LISTENER} normalizes the direct acoustic path between the listener and loudspeaker at $\VEC{u}_n$ to be its on-axis response $S_A(\nu, 0)$ at the normalized coordinate $\VEC{v}_n$.
}
\label{FIG:PT:PRETRANSFORM}
\end{center}
\end{figure}

Consider the following decomposition of the loudspeaker transfer function $S(\nu, \theta) = S_E(\nu)  S_A(\nu, \theta) $ into acoustical and electrical domain transfer functions $S_A(\nu, \theta)$ and $S_E(\nu)$ respectively. A filter with frequency response $G_n(\nu)$ that normalizes \eqref{EQ:PT:ANECHOIC_TF} to the loudspeaker's on-axis response  $H_n(\nu, \VEC{0}) G_n(\nu)  = S(\nu, 0)$ is given by
\begin{equation}
\begin{split}
G_n(\nu)   = Q_n(\nu)   \NORM{\VEC{u}_n } e^{ j \kappa \NORM{\VEC{u}_n}}, \quad 
Q_n(\nu)   = \frac{S_A(\nu, 0)}{ S_A \PR{\nu,  \VEC{\bar{\theta}}_n } },
\end{split}
\label{EQ:PT:ANECHOIC_TF_LISTENER}
\raisetag{8.5ex}
\end{equation}
where $ \VEC{\bar{\theta}}_n = \theta_n(\VEC{0})$ is the normalization angle between the loudspeaker's orientation and the listener. The electrical domain term $S_E(\nu)$ cancels within the quotient $Q_n(\nu)$ in \eqref{EQ:PT:ANECHOIC_TF_LISTENER}, thereby negating prior signal processing in loudspeaker playback. $Q_n(\nu)$ is therefore the acoustic relative-transfer-function between loudspeaker's axial and listener-direction acoustic responses. Moreover, if $S_A(\nu, \theta)$ share a common acoustic delay and the remainder is minimum-phase for bounded $\theta$ that define a listening window, then $Q_n(\nu)$ must also be minimum-phase. Thus, the normalized transfer function $G_n(\nu)$ compensates for both loudspeakers' orientation and distance relative to the listener as shown in Fig. \ref{FIG:PT:PRETRANSFORM}.

In practice, we can find the rational function approximation \citep{sanathanan2003transfer, lai2010minimax} to $Q_n(\nu)$, expressed in terms of minimum-phase $\FIELD{M}_n(\nu)$ and all-pass $\FIELD{A}_n(\nu)$ transfer functions given by
\begin{equation}
\begin{split}
Q_n(\nu) \approx \FIELD{M}_n(\nu) \FIELD{A}_n(\nu), \quad 
\FIELD{A}_n(\nu) = \bar{\FIELD{A}}_n(\nu) \ddot{\FIELD{A}}_n(\nu),
\end{split}
\label{EQ:PT:TF_MP_AP}
\end{equation}
where $\bar{\FIELD{A}}_n(\nu)$ and $\ddot{\FIELD{A}}_n(\nu)$ are all-pass transfer functions belonging to stable and unstable components respectively. The unstable all-pass $\ddot{\FIELD{A}}_n(\nu)$ contains the reciprocal poles and zeros of the Padé approximant outside the complex unit-circle, and is ideally empty or low-order for $\theta$ in the listening window. We can realize a causal-stable filter-response $G_n(\nu)$ for an all-passed loudspeaker transfer function in \eqref{EQ:PT:ANECHOIC_TF_LISTENER} as follows:
\begin{equation}
\begin{split}
H_n(\nu, \VEC{0}) G_n(\nu)  = S(\nu, 0)  \frac{  e^{ \minus j \kappa d} }{\ddot{\FIELD{A}}_{lcm}(\nu)}  \quad \Rightarrow \quad \\
G_n(\nu)   =    \FIELD{M}_n(\nu)    \bar{\FIELD{A}}_n(\nu)   \frac{ \ddot{\FIELD{A}}_n(\nu)  }{  \ddot{\FIELD{A}}_{lcm}(\nu) } \NORM{\VEC{u}_n }   e^{ j \kappa \NORM{\VEC{u}_n  \minus d }}, \\
\end{split}
\label{EQ:PT:CAUSAL_FILT}
\end{equation}
where $d = \max_{1 \leq n \leq N} \NORM{\VEC{u}_n}$ is the furthest loudspeaker distance, and $\ddot{\FIELD{A}}_{lcm}(\nu)$ is the transfer function of the set of least common multiple (LCM) reciprocal poles and zeros across the unstable all-passes $\CR{ \ddot{\FIELD{A}}_1 (\nu) , \hdots,  \ddot{\FIELD{A}}_N (\nu) }$. In the $z$-domain, we can therefore express the all-pass and LCM transfer functions as follows: 
\begin{equation}
\begin{split}
\ddot{\FIELD{A}}_{n}(z) & =   \prod_{p \in P_n}  \PR{\frac{1-p^* z}{1-pz^{\minus 1}}}^{k_{pn} }, \quad 
P_n = \CR{p_{1n},  \hdots, p_{M_n n}},  \\
\ddot{\FIELD{A}}_{lcm}(z) & = \prod_{p \in P }  \PR{\frac{1-p^* z}{1-pz^{\minus 1}}}^{\max\limits_{1 \leq n \leq N} k_{pn} }, \quad
P = \cup_{n=1}^N P_n,
\end{split}
\label{EQ:PT:AP_LCM}
\raisetag{7.5ex}
\end{equation}
where $p^*$ is the conjugate transpose, and $P_n$ is the set of unique poles and $k_{pn}$ is the multiplicity of pole $p$ for the $n^{th}$ loudspeaker. By taking the maximum multiplicity for each unique and unstable pole across all $\ddot{\FIELD{A}}_{n}(z)$, and dividing by the subsequent LCM $\ddot{\FIELD{A}}_{lcm}(z)$, the unstable poles in $\ddot{\FIELD{A}}_{n}(z)$ cancel and the remaining all-pass adds minimal additional group-delay in $G_n(\nu)$. The filtered loudspeakers' direct paths are thus matched with a common all-passed on-axis response. Lastly, we gain the loudspeaker filter $G_n(\nu)$ to match the expected acoustic power at a common distance $D$, such as the median of all loudspeakers-to-listener distances, via the following room acoustic attenuation model: 

Let us consider the inverse-distance law $\rho_{DP}(r) = \bar{\rho} r^{\minus 2}$ for the attenuation of the direct acoustic path response's nominal power $\bar{\rho}$ at distance $r$ from a loudspeaker. In a room environment, let $\rho_{IP}(r)$ be the total power of indirect acoustic paths at distance $r$. We can model the ratio of the direct-to-indirect acoustic path's power at $r$ and total power as follows:
\begin{equation}
\begin{split}
\frac{\rho_{DP}(r)}{\rho_{IP}(r)} & = \PR{\frac{d_c}{r}}^{2\beta}, \quad 
\beta = 10^{\frac{\gamma \textrm{ dB/dd}}{10} }, \quad \textrm{Attenuation rate}
\\ 
\rho(r) & =  \rho_{DP}(r) + \rho_{IP}(r)   = \bar{\rho} r^{\minus 2} \PR{1 + \PR{\frac{d_c}{r}}^{2 \beta} },
\end{split}
\raisetag{5ex}
\label{EQ:PT:RAT}
\end{equation}
where $d_c$ is the so-called \textit{critical distance} (meters) where the direct and indirect acoustic powers are equivalent, and $\beta$ a decay-rate parameterized by $\gamma$ decibels (dB) per double-distance (dd); typical $\gamma \in \CR{0, -3}$ and $0.5 \leq d_c \leq 1.5$ span idealized concert-hall to small-room spaces \citep{toole2006loudspeakers}. Normalizing the power at distance $r$ to $D$ therefore follows
\begin{equation}
\begin{split}
F(r, \, D, \, d_c) = \sqrt{\frac{\rho(D) }{\rho(r) }}  =  \frac{r}{D} \sqrt{\frac{d_c^{2\beta} + D^{2\beta} }{d_c^{2 \beta } + r^{2\beta} }}, 
\end{split}
\label{EQ:PT:DIST}
\end{equation}
whereby substituting $\NORM{\VEC{u}_n }$  with $F(\NORM{\VEC{u}_n }, \, D, \, d_c)$ in \eqref{EQ:PT:CAUSAL_FILT} compensates for loudspeaker distances to the listener in a room.

\textbf{Model Uncertainty for Non-stationary Targets:}
In instances where the listener's location changes over time or require online estimation, we normalize the loudspeaker via the mean listener distance $\frac{1}{T} \int_{0}^{T} \NORM{\VEC{u}_n(t)} dt$, and treat the normalization angle  $\VEC{\bar{\theta}}_n$ relative to the loudspeaker orientation $\VEC{o}_n$ in \eqref{EQ:PT:ANECHOIC_TF_LISTENER} as a random variable. The target transfer function $G_n(\nu)$ and quotient term $Q_n(\nu)$ are re-defined to minimize the expected squared-differences between the anechoic responses $S_A(\nu, \theta)$ sampled over axial-centered and loudspeaker-listener centered circular probability distribution functions (PDFs) $f_0(\theta)$ and $f_n(\theta)$, $\forall 1 \leq n \leq N$ respectively; circular PDFs satisfy $f(\theta) = f(\theta + 2\pi k)$, $\forall k \in \FIELD{Z}$. We present two acoustic averages:
\begin{equation}
\begin{split}
\bar{S}_A(\nu, f(\theta)) & = \EXPECTATION{ }{S_A(\nu, \theta) } = \int S_A(\nu, \theta) f(\theta) d \theta, \\
\hat{S}_A(\nu, f(\theta)) & = \EXPECTATION{ }{\ABS{S_A(\nu, \theta)}^2 } = \int \ABS{S_A(\nu, \theta)}^2 f(\theta) d \theta,
\end{split}
\label{EQ:PT:AVG}
\raisetag{9.5ex}
\end{equation}
where $\bar{S}_A(\nu, f(\theta) )$ and $\hat{S}_A(\nu, f(\theta) )$ are spatial windowed averages of the acoustic response and power respectively; axial window response average $\bar{S}_A(\nu, f_0(\theta) )$ and power average $\hat{S}_A(\nu, f_0(\theta) )$ sample from the $f_0(\theta)$ distribution. The modified quotient term $Q_n(\nu)$ in \eqref{EQ:PT:ANECHOIC_TF_LISTENER} is replaced with the weighted least-squares minimizer of $\argmin_{X} \int \ABS{ S_A(\nu, \theta)  X -   \bar{S}_A(\nu, f_0(\theta))   }^2 f_{n}(\theta) d \theta$ given by
\begin{equation}
\begin{split}
\bar{Q}_n(\nu) =  \bar{S}_A(\nu, f_0(\theta) ) \frac{ \bar{S}_A^*(\nu, f_n(\theta) )  }{ \hat{S}_A(\nu, f_n(\theta) )  },
\end{split}
\label{EQ:PT:MIN}
\raisetag{5ex}
\end{equation}
where $\bar{S}_A^*(\nu, f_n(\theta) )$ is the conjugate transpose, and  $\bar{Q}_n(\nu)$ accounts for both amplitude and phase differences in the averaged responses. The analogous quotient for the spatial windowed acoustic power average follows
\begin{equation}
\begin{split}
\hat{Q}_n(\nu) & =  \sqrt{\frac{ \hat{S}_A(\nu, f_0(\theta)) }{  \hat{S}_A(\nu, f_n(\theta)) }},
\end{split}
\label{EQ:PT:RAT}
\end{equation}
where $\hat{Q}_n(\nu)$ has zero-phase and therefore compensates for only the amplitude. Both quotients can be efficiently evaluated if $f_0(\theta)$, $f_n(\theta)$ are both uni-modal and smooth over azimuth, have expansions along a common orthogonal basis with $S_A(\nu, \theta)$, and follow the contours of a listening window.

\begin{figure}[h]
\begin{center}
  {\includegraphics[width=0.95\columnwidth]{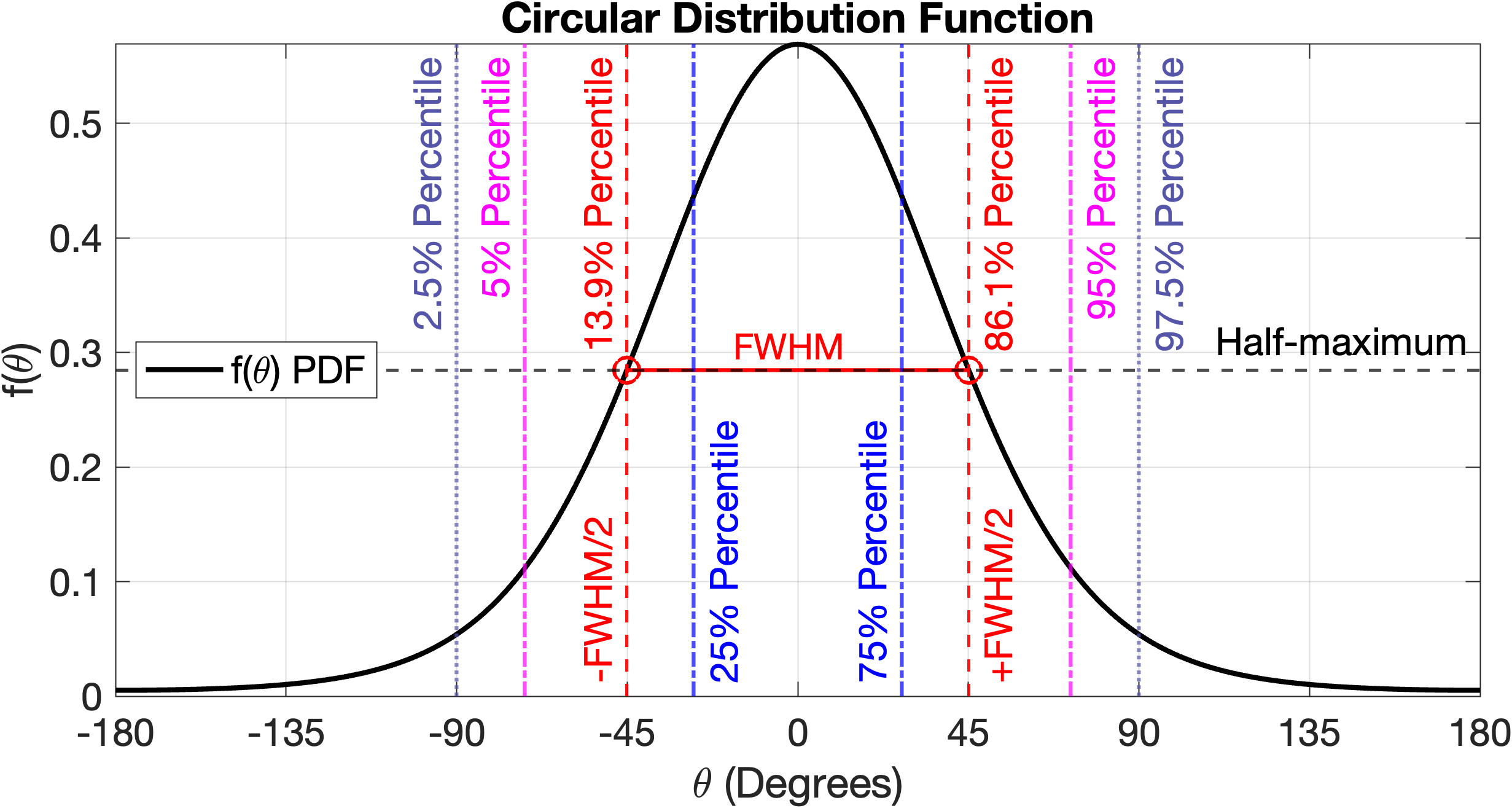}}
\caption{Circular distribution prior (FWHM $90.2^{\circ}$) contains $\CR{50, 90, 95}\%$ of  normalization angles within $\ABS{\theta} \leq \CR{27.4,\, 72,\, 90}^{\circ}$ of the mean angle. }
\label{FIG:PT:WIN}
\end{center}
\end{figure}

Let us consider the circular distribution $f(\theta)$ defined by the squared-exponential of the chordal distance $d(\theta)$ on a unit-disk, which along with $S_A(\nu)$ has a series-expansion over the Legendre polynomials \citep{LUO_2021_SPH}, and normalized over the domain of all azimuth angles $-\pi \leq \theta \leq \pi$:
\begin{equation}
\begin{split}
f(\theta) & =  \frac{e^{\frac{\minus d^2(\theta \minus \mu)}{2 \ell^2} }}{ 2 \pi e^{\minus \ell^{\minus 2}} J_0 (j \ell^{\minus 2}) }, \quad 
d(\theta)  = 2 \sin \PR{\frac{\MOD{\theta}{2 \pi}}{2}},
\end{split}
\label{EQ:PT:PDF}
\raisetag{7.5ex}
\end{equation}
where $J_0$ is the Bessel function of the first kind, $\mu$ is the mean azimuth, and $\ell$ is the dispersion. The function is symmetric w.r.t. its maximum $f(\mu)$ and minimum $f(\mu \pm \pi)$, infinitely differentiable in all azimuths, and its percentiles computable via series expansion in appendix \eqref{EQ:APPENDIX:CD:INT}. Large dispersion $\ell$ gives a uniform distribution as $\lim_{\ell \rightarrow \infty} f(\theta) =  (2 \pi)^{\minus 1}$; small dispersion gives the dirac distribution as  $\lim_{\ell \rightarrow 0} f(\theta - \mu) =  \delta$. We can bound the dispersion via design parameters characterizing a listening window's peak such as the full-width at half-maximum (FWHM) measure:
\begin{equation}
\begin{split}
\frac{f(\mu)}{2} & = f\PR{\mu \pm \frac{\textrm{FWHM}}{2}}, \quad
0 \leq \textrm{FWMH} \leq 2\pi, \\
\ell & = \frac{2 \sin \PR{ \frac{\textrm{FWMH}}{4} } }{ \sqrt{2 \ln(2) }} \,\,\, \Rightarrow \,\,\,
0 \leq \ell \leq \sqrt{2 / \ln(2)},
\end{split}
\label{EQ:PT:FWHM}
\end{equation}
which defines the angular width where $f(\theta)$ spans half its maximum amplitude as shown in Fig. \ref{FIG:PT:WIN}. At the upper-limit FWHM $360^{\circ}$, $f(\theta)$ contains $\CR{60.9,\, 33.2,\, 22.5}\%$ of its mass within the frontal intervals $\ABS{\theta - \mu} \leq \CR{90,\, 45,\, 30}^{\circ}$ respectively. For tighter FWHM  $\leq 90.22^{\circ}$ bounds, $f(\theta)$ contains the $95\%$ confidence interval in the half-space $\ABS{\theta -\mu} \leq 90^{\circ}$ of its mean azimuth $\mu$. For the axial-centered PDF in \eqref{EQ:PT:AVG}, we set the window's FWHM to $60^{\circ}$ where $f_0(\theta) = f(\theta \, | \, \mu = 0, \ell = 0.4396)$. We now proceed with online adaptation of the normalization angles $\VEC{\bar{\theta}}_n$ over time.

Suppose we have measured a normalization angle $\bar{\theta}$ belonging to the $n^{th}$ loudspeaker with known measurement dispersion $\bar{\ell}$ such that the likelihood function $f(\bar{\theta} \, | \, \mu = \VEC{\bar{\theta}}_n,  \ell = \bar{\ell})$ follows the squared-exponential chordal function in \eqref{EQ:PT:PDF}. Let the unknown normalization angle $\VEC{\bar{\theta}}_n$ of the $n^{th}$ loudspeaker have a squared-exponential chordal prior-distribution $f(\VEC{\bar{\theta}}_n | \, \mu =  \mu_n, \ell = \ell_n)$  with initial hyperparameters $\mu_n = 0$ mean azimuth and $\ell = \sqrt{2 / \ln(2)}$ maximum dispersion. The posterior normalization angle therefore has a conjugate distribution with hyperparameters following appendix \eqref{EQ:APPENDIX:CD:PROD_CD}. Over multiple time-steps $t$, the likelihood, prior, and posterior functions across measured angles $\VEC{\bar{\theta}}^{\CR{t}}_n$ with dispersion $\bar{\ell}_n^{\CR{t}}$ are given by
\begin{equation}
\begin{split}
L\PR{\VEC{\bar{\theta}}_n \, | \, \VEC{\bar{\theta}}^{\CR{t}}_n } & =  f \PR{ \VEC{\bar{\theta}}^{\CR{t}}_{n}  \, | \, \mu = \VEC{\bar{\theta}}_n, \ell = \bar{\ell}_n^{\CR{t}} }, \quad \textrm{Likelihood} \\
P(\VEC{\bar{\theta}}_n ) & = f \PR{ \VEC{\bar{\theta}}_n \, | \, \mu =  \mu_n^{\CR{t \minus 1}},  \, \ell = \ell_n^{\CR{t \minus 1}} }, \quad \textrm{Prior} \\
P\PR{\VEC{\bar{\theta}}_n \, | \, \VEC{\bar{\theta}}^{\CR{t}}_n  } &  \propto L\PR{\VEC{\bar{\theta}}_n \, | \, \VEC{\bar{\theta}}^{\CR{t}}_n } P(\VEC{\bar{\theta}}_n), \qquad \textrm{Posterior}
\end{split}
\label{EQ:PT:PDF_STAT}
\raisetag{3.5ex}
\end{equation}
where the reported normalization angle $\VEC{\bar{\theta}}^{\CR{t}}_n$ is a point-estimate taken within a measurement session, and the dispersion $\bar{\ell}_n^{\CR{t}}$ is proportional to the point-estimate's confidence interval. Both quantities can vary over time as the listener's location may change between sessions (e.g. different seating), and measured under different noise conditions. The initial hyperparameters for mean $\mu_n^{\CR{0}} = 0$  and dispersion $\ell_n^{\CR{0}} = 0.6515$ (FWHM $90.22^{\circ}$) are informative as loudspeakers generally orient towards the intended listening area. The posterior estimate of $\VEC{\bar{\theta}}_n$  follows Bayes' theorem, where the current mean $\mu_n^{\CR{t}}$ and dispersion $\ell_n^{\CR{t}}$  hyperparameters are updated from the measurement terms $\VEC{\bar{\theta}}^{\CR{t}}_n, \bar{\ell}_n^{\CR{t}}$ in the likelihood function and the previous hyperparameters $\mu_n^{\CR{t-1}}, \ell_n^{\CR{t-1}}$ via appendix \eqref{EQ:APPENDIX:CD:PROD_CD_PRM}. Lastly, the normalization filter's quotient terms \eqref{EQ:PT:MIN}, \eqref{EQ:PT:RAT} are updated for PDF $f_n(\theta) = f(\theta \, | \, \mu = \mu_n^{\CR{t}}, \ell = \ell_n^{\CR{t}} )$, and the filters $G_n(\nu)$ are re-computed. Let us step-through the following example:

\begin{figure*}[h]
\begin{center}
  {\includegraphics[width=0.3375\textwidth]{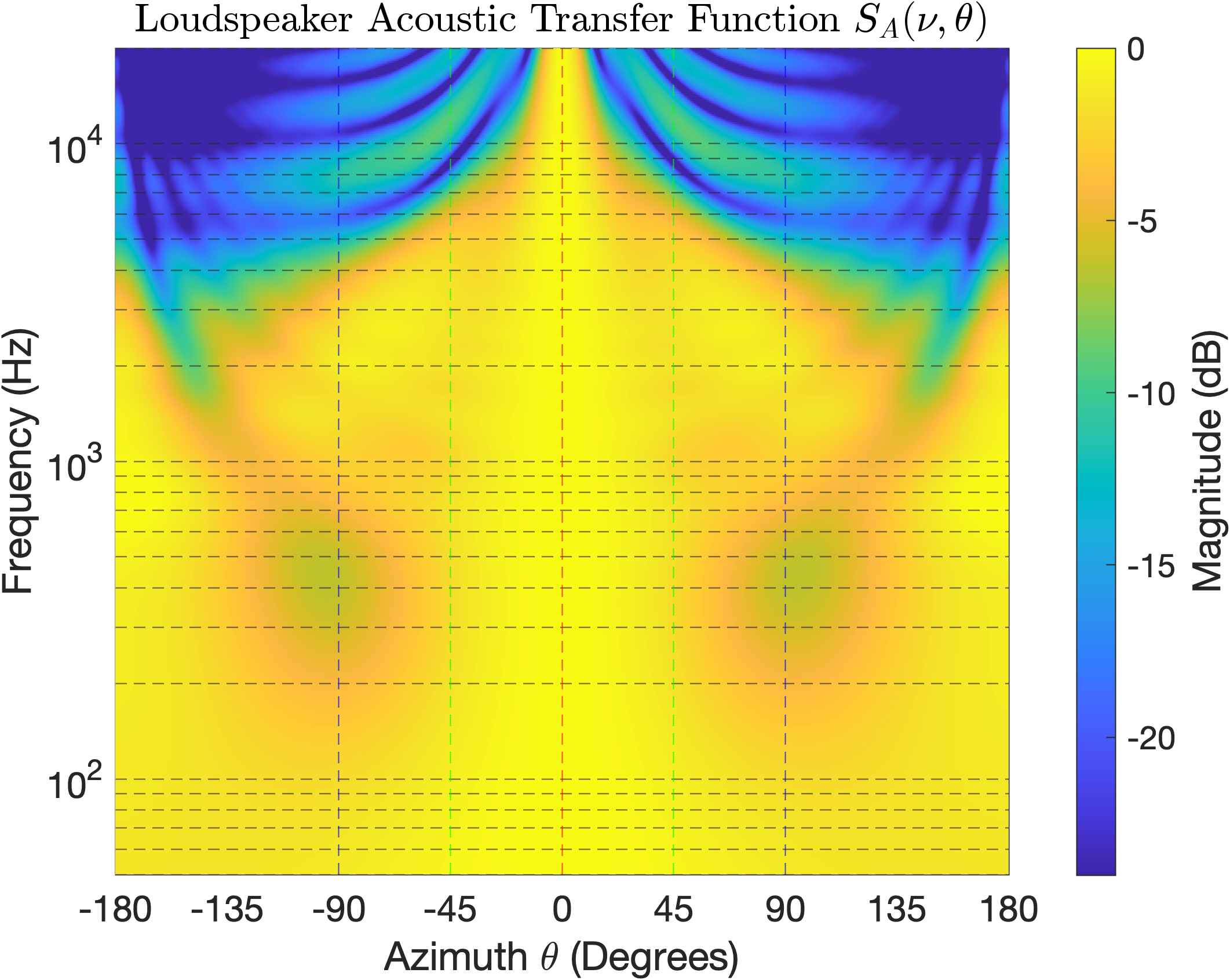}}
    {\includegraphics[width=0.325\textwidth]{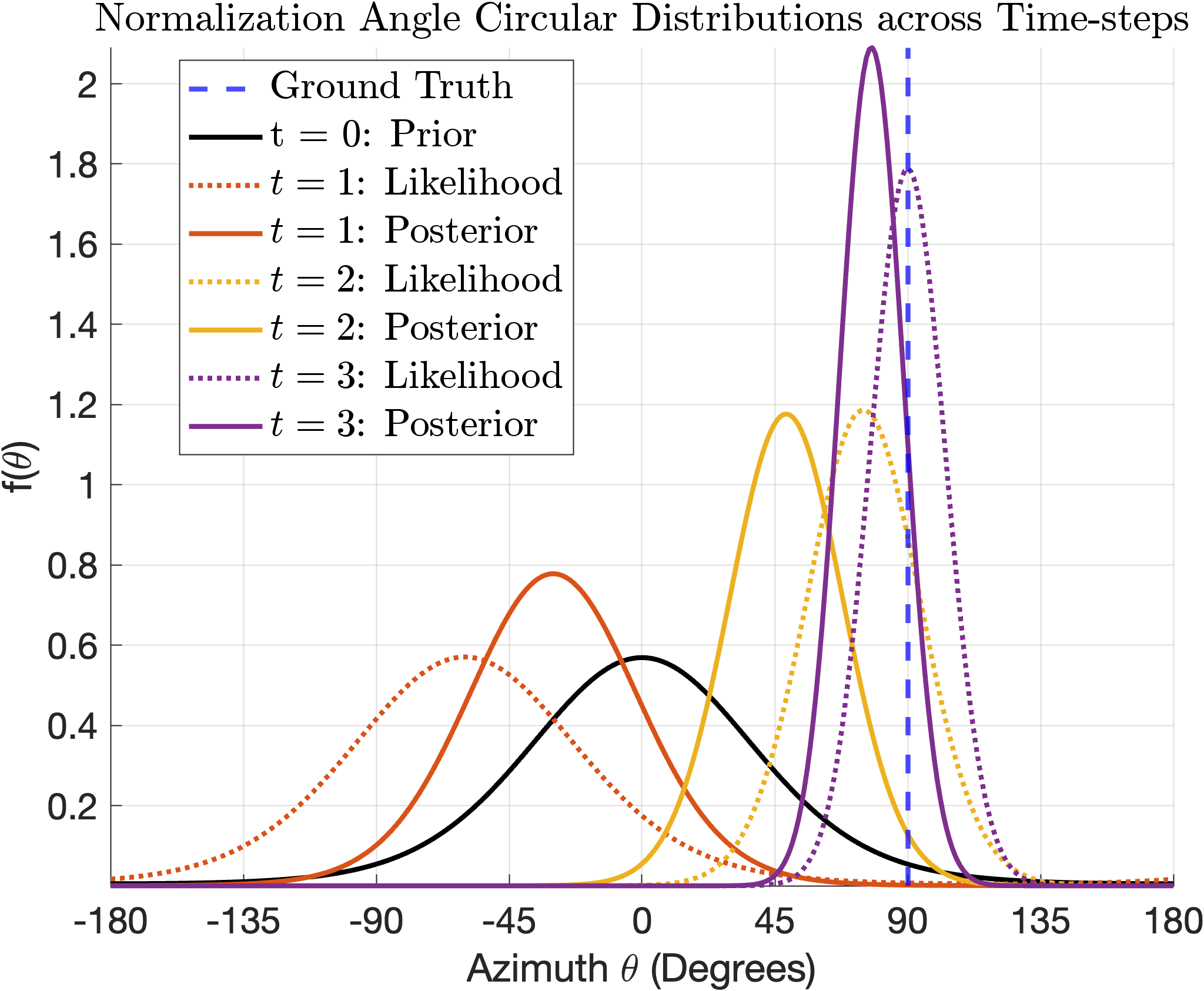}}
      {\includegraphics[width=0.325\textwidth]{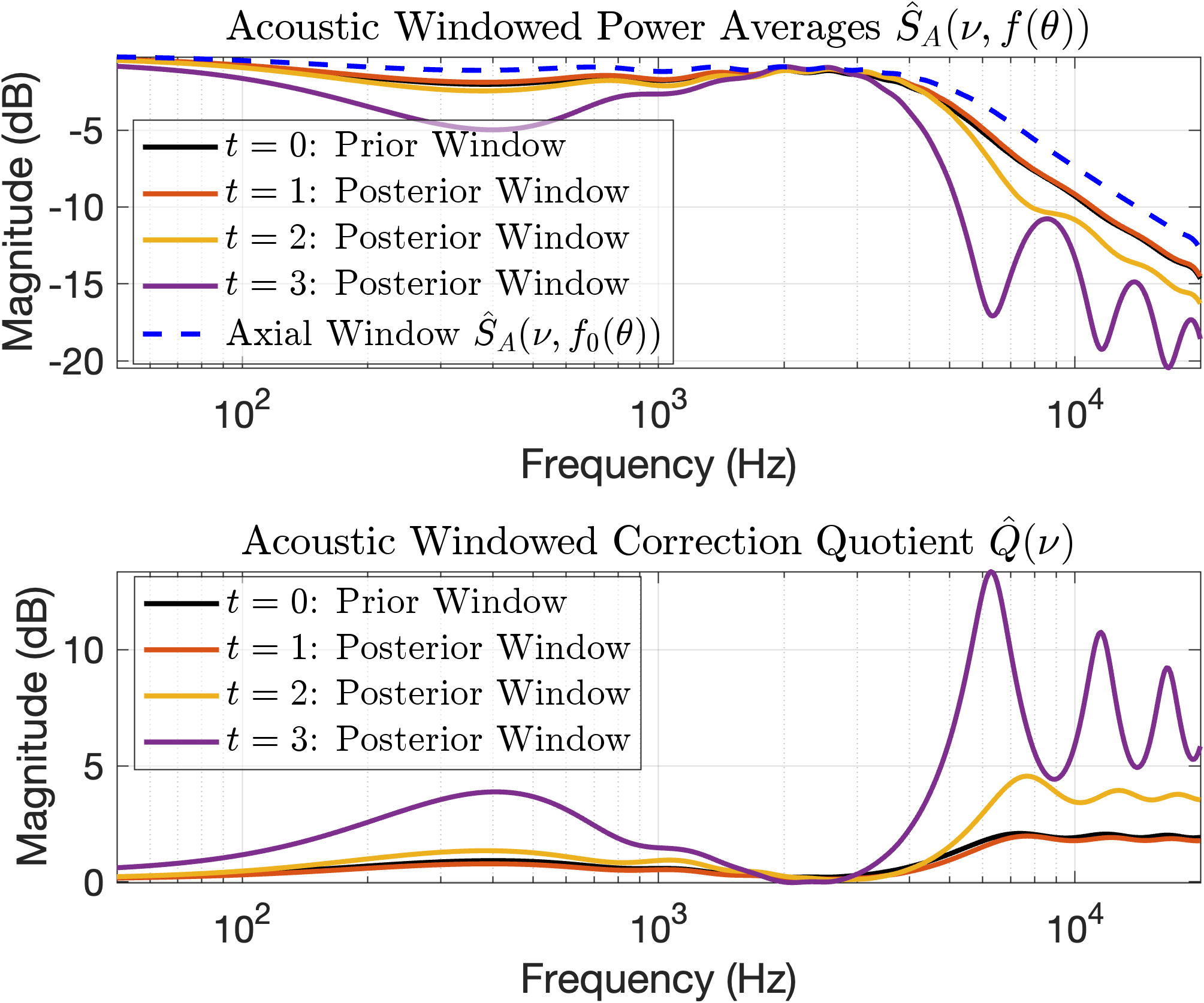}}
\caption{We equalize a sample loudspeaker with acoustic responses over the horizontal plane (left) between Bayesian estimates of the normalization angle $\VEC{\bar{\theta}}$ in \eqref{EQ:PT:PDF_STAT} (center) and the axial windowed power average. The acoustic power averages (right) over the posterior circular distribution windows $f(\theta \, | \, \mu = \mu^{\CR{t}}, \ell = \ell^{\CR{t}} )$ update across time-steps to yield a sequence of quotient correction targets in \eqref{EQ:PT:RAT}. }
\label{FIG:PT:BAYES}
\end{center}
\end{figure*}

Consider the sample loudspeaker responses and sequence of estimated normalization angles in Fig. \ref{FIG:PT:BAYES} where the listener is $90^{\circ}$ offset the loudspeaker axis in azimuth. At $t=0$ prior to any measurements, the normalization angle assumes a circular distribution centered on the loudspeaker axis $\mu = 0$ with wide dispersion FWHM $90.22^{\circ}$. The first estimate $\VEC{\bar{\theta}}^{\CR{1}} = -60^{\circ}$ is inaccurate with high dispersion FWHM $90^{\circ}$ as shown in the $t=1$ likelihood. Although the posterior shifts its mean halfway between the prior's mean and estimated angle, the dispersion remains high, which gives a similar acoustic windowed power average and correction quotient to that of the prior. The second estimate  $\VEC{\bar{\theta}}^{\CR{2}} = 75^{\circ}$ is more accurate with lower dispersion FWHM $45^{\circ}$. The resulting posterior shifts much closer towards the estimate at much reduced dispersion, which distinguishes its windowed power average and correction quotient from the prior. The final and most accurate estimate $\VEC{\bar{\theta}}^{\CR{3}} = 90^{\circ}$ with lowest dispersion FWHM $30^{\circ}$ yields a sharp posterior near the true normalization angle, which induces comb-filter patterns in the correction quotient due to lobbing in the loudspeaker's anechoic response in azimuth. Therefore in practice, we avoid equalizing to direct acoustic-paths by enforcing a lower-bound dispersion FWHM $45^{\circ}$ for circular distributions $f_n(\theta)$ when computing the correction quotients $\hat{Q}_n(\nu)$.

\section{Loudspeaker Panning Optimization}
\label{SEC:DC}

Let $R_n(\nu, \VEC{r}) =  H_n(\nu, \VEC{r}) G_n(\nu)$ be the acoustic response at frequency $\nu$ and coordinate $\VEC{r}$ of the $n^{th}$ normalized loudspeaker in \eqref{EQ:PT:CAUSAL_FILT}, and the overall response of the normalized loudspeaker array follows
\begin{equation}
\begin{split}
Y(\nu, \VEC{r}) = \sum_{n=1}^N R_n(\nu, \VEC{r}) X_n(\nu),
\end{split}
\label{EQ:DC:ARRAY}
\end{equation}
where $X_n(\nu)$ is the transfer function of the array's weights belonging to the $n^{th}$ loudspeaker. For normalized loudspeaker panning, we constrain $X_n(\nu)$ to have a common phase-component (e.g. delay or all-pass) across loudspeakers and solve for the unknown magnitude components $x_n(\nu) = \ABS{X_n(\nu)}$, which are subject to frequency-dependent spatial-electrical-acoustic domain constraints. The magnitude components at frequency $\nu$ are therefore expressed as  a vector of panning gains  $\VEC{x} = \BK{x_1, \hdots x_N}^T \in \FIELD{R}^{N \times 1}$, whereby we omit the frequency $\nu$ specification for simplifying notation. 
Further simplifications following the loudspeaker normalization are possible when specifying domain-specific constraints. Loudspeaker coordinates reduce to their unit-directions in the spatial domain given by 
\begin{equation}
\begin{split}
\MAT{V} = \BK{\VEC{v}_1, \hdots, \VEC{v}_N} \in \FIELD{R}^{2 x N}, \quad 
\VEC{v}_n = \frac{\VEC{u}_n}{ \NORM{\VEC{u}_n}}.
\end{split}
\label{EQ:DC:DIR}
\end{equation}
The normalization filter's electrical gain $\ABS{G_n(\nu)}$ bounds the electrical headroom in the electrical domain. The normalized loudspeaker acoustic responses in \eqref{EQ:PT:CAUSAL_FILT} are matched at the listener's location in the acoustical domain. 

\textbf{Spatial Panning Constraints:}
The vector-base amplitude panning with slack (VBAPS) constraint is given by
\begin{equation}
\begin{split}
\MAT{V} \VEC{x} = \lambda \VEC{s}, \quad 
\VEC{x} \geq \VEC{0}, \quad 
\lambda \geq 0,
\end{split}
\label{EQ:DC:SDC:VBAPS}
\end{equation}
where the panning gains $\VEC{x}$ are non-negative as to preserve the relative-phase between loudspeaker pairs, and constrain the weighted average of the loudspeaker directions $\MAT{V}$ to coincide with the target steering unit-direction $\VEC{s} \in \FIELD{R}^{2 \times 1}$ upto non-negative scale given by the slack-variable $\lambda$. The latter is an augmented variable for both scaling the target unit-direction $\VEC{s}$ to lie in equality with the panning direction $\MAT{V}\VEC{x}$ as shown in Fig. \ref{FIG:DC:SDC:CONSTR}, and to accommodate constraints placed on $\VEC{x}$ from other domains. The feasible steering and panning directions, and panning gains are therefore constrained as follows:

\begin{figure}[h]
\begin{center}
  {\includegraphics[width=0.49\columnwidth]{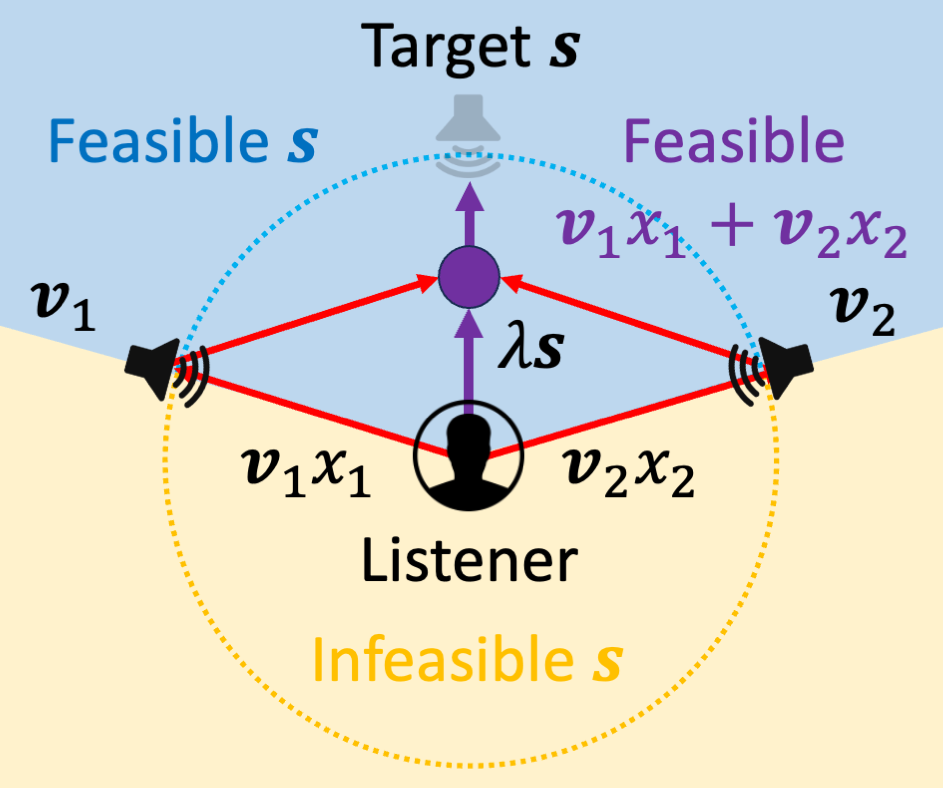}}
   {\includegraphics[width=0.49\columnwidth]{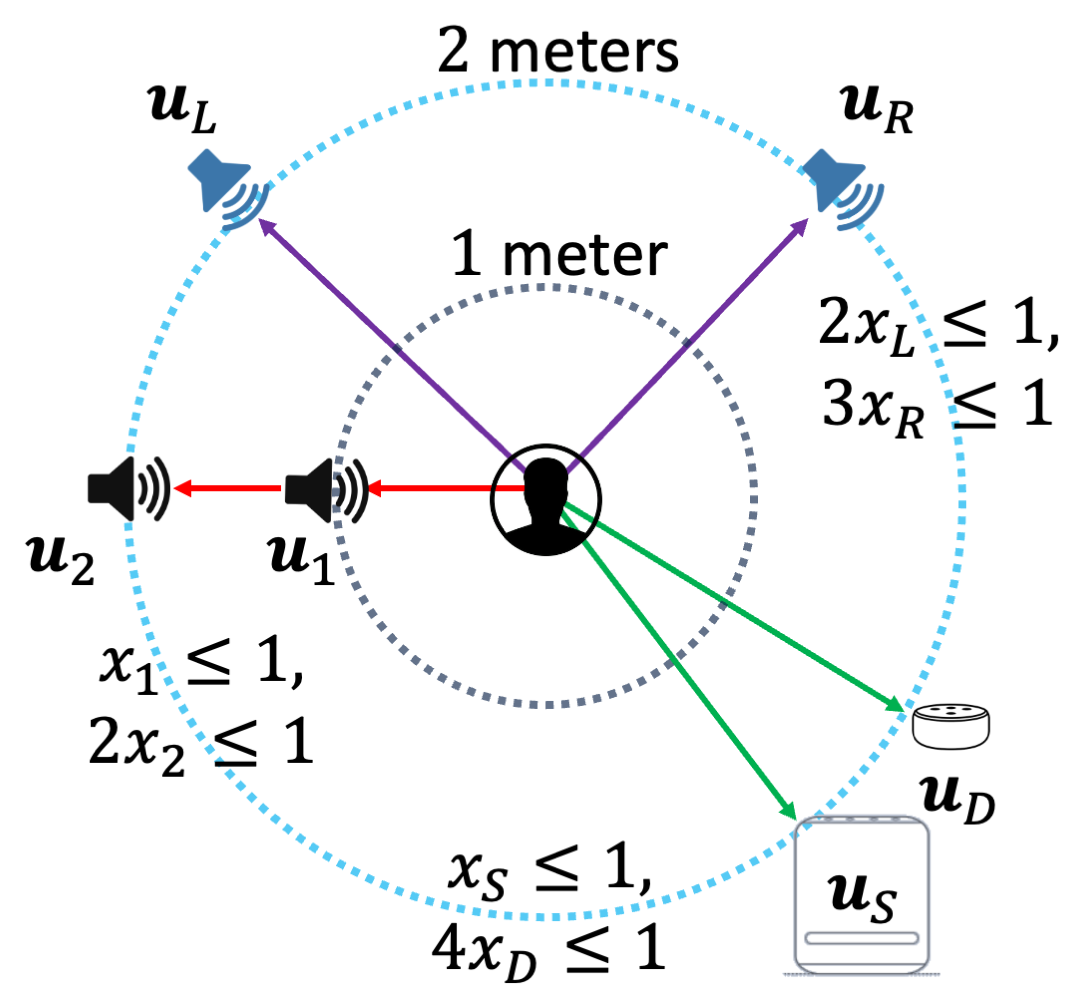}}
\caption{VBAPS (left) constrains the feasible steering direction $\VEC{s}$ to lie between the minor-arc of the loudspeaker pair coordinates $\VEC{x}_L, \VEC{x}_R$. Sample voltage constraints (right) are proportional to differences in loudspeaker-to-listener distance, orientation, and selection.}
\label{FIG:DC:SDC:CONSTR}
\end{center}
\end{figure}

Consider a set of $N$ loudspeakers and panning gains satisfying \eqref{EQ:DC:SDC:VBAPS}. The set of feasible steering unit-directions $\VEC{s}$ must lie in the union of minor-arcs between all pairwise loudspeaker unit-directions. Conversely, steering directions are infeasible along the major-arc of a single loudspeaker-pair $N=2$ as shown in Fig. \ref{FIG:DC:SDC:CONSTR}. For $N>2$ loudspeakers, the feasible $\VEC{s}$ are all of $\FIELD{R}^{2}$ iff there exist a set of three loudspeakers where the negative direction of each loudspeaker lies between the minor-arc of the other two loudspeaker directions. The panning direction $\MAT{V}\VEC{x}$ is therefore constrained to be in the set of $\lambda$-scaled feasible unit-directions $\VEC{s}$. We now introduce several evaluation metrics or objectives w.r.t. $\lambda$.

Let us define \textit{panning sensitivity} by the acoustic-path distance ratio of the panning direction and the summation of component panning gained loudspeaker directions given by
\begin{equation}
\begin{split}
\FIELD{S}(\MAT{V}, \VEC{x}, \VEC{s}) = \frac{ \NORM{\MAT{V}\VEC{x} } }{ \sum_{n=1}^N \NORM{\VEC{v}_n x_n} } =  \frac{\NORM{ \lambda \VEC{s} }}{ \sum_{n=1}^N x_n } = \frac{\lambda}{ \VEC{x}^T \VEC{1}},
\end{split}
\label{EQ:DC:SDC:SENSITIVITY}
\end{equation}
which has bounds $0 < \FIELD{S}(\MAT{V}, \VEC{x}, \VEC{s})  \leq 1$. Sensitivity is maximal iff non-zero panning gains belong to loudspeakers with directions coincident to the steering direction, large if panning gains disproportionately allocate to loudspeakers with directions closer to the steering direction, and minimal when panning gains allocate to loudspeakers with directions that sum to zero. Panning sensitivity therefore gives a similarity measure between panned and discrete sound-sources in the direction of $\VEC{s}$. This contrasts with cross-domain measures of \textit{panning efficiency}, which evaluates the power ratios between panning direction and electric or acoustic gain as follows:
\begin{equation}
\begin{split}
\FIELD{F}(\MAT{K}, \MAT{V}, \VEC{x}) =  \frac{\VEC{x}^T \MAT{V}^T \MAT{V} \VEC{x}}{\VEC{x}^T \MAT{K} \VEC{x}} =  \lambda^2 \NORM{\MAT{K}^{\frac{1}{2}} \VEC{x}}^{\minus 2},
\end{split}
\label{EQ:DC:SDC:EFFICIENCY}
\end{equation}
where $\MAT{K} \in \FIELD{C}^{N \times N}$ is a domain-dependent covariance matrix (identity for electrical, model dependent for acoustical).  For the electrical domain where $\MAT{K} = \MAT{I}$, the maximum efficiency is $N$ for loudspeakers with directions coincident to the steering direction and uniform panning gains $\VEC{x} =  N^{\minus 1}\VEC{1}$. For the acoustic domain, the maximum efficiency is the largest generalized eigenvalues between $\MAT{V}^T \MAT{V}$ and $\MAT{K}$. Thus, higher panning efficiency is realized via more uniformly distributed panning gains across loudspeakers, whereas high panning sensitivity follows sparsely distributed panning gains.

\textbf{Electrical Headroom Constraints:}
The electrical-power headroom of normalized loudspeakers decreases in proportion to the normalization filter power responses $\ABS{G_n(\nu)}^2$. Under non-negative panning constraint, the quadratic electrical-power constraint are linearized as follows:
\begin{equation}
\begin{split}
x_n x_n^*  & \leq \ABS{G_n(\nu)}^{\minus 2}, \quad 
x_n \geq 0, \quad
 \Rightarrow \quad 
\VEC{0} \leq \VEC{x} \leq \VEC{\tau},
\end{split}
\label{EQ:DC:EHC:ELECTRICAL_HEADROOM}
\end{equation}
where $\VEC{\tau} = \BK{\ABS{G_1(\nu)}^{\minus 1}, \hdots, \ABS{G_N(\nu)}^{\minus 1}}^T \in \FIELD{R}^{N \times 1}_{\geq 0}$ is a vector containing the digital headroom per loudspeaker that bounds the feasible space of panning gains to the upper box-orthant. We give several examples of voltage headroom consumed by normalization in Fig. \ref{FIG:DC:SDC:CONSTR}. Doubling the loudspeaker $\VEC{u}_1$'s distance to the listener to that of $\VEC{u}_2$ halves the voltage headroom. Re-orienting the loudspeaker $\VEC{u}_R$ to face the listener at $\VEC{u}_L$ lowers high-frequency headroom. Equalizing the mid-range loudspeaker at $\VEC{u}_D$ to match the full-range loudspeaker at $\VEC{u}_S$ decreases the low-frequency headroom.

\textbf{Acoustical Power Constraints:}
The acoustic covariances between the normalized loudspeaker transfer functions $R_n(\nu, \VEC{r})$ in \eqref{EQ:DC:ARRAY}, over coordinates $\VEC{r}$ in the listening area, specify quadratic power constraints in equality to the acoustic power target $\rho$ as follows:
\begin{equation}
\begin{split}
\VEC{x}^T \MAT{K} \VEC{x} & =  \rho, \quad 
K_{ij} \approx  \EXPECTATION{\VEC{r} \sim g(\VEC{r}) }{R_{i}(\nu, \VEC{r} ) R_{j}^*(\nu, \VEC{r})  },
\end{split}
\label{EQ:DC:ALC:ACOUSTIC_POW}
\end{equation}
whereby $\VEC{r}$ is sampled from a disc of radius $\tau_r$ with a truncated uniform PDF $g(\VEC{r}) = \frac{1}{\pi \tau_r^2}, \forall \, \NORM{\VEC{r}} \leq \tau_r$, and $0$ otherwise. For loudspeaker transfer functions in the far-field,  spherical-waves can be approximated by plane-waves which give the  acoustic covariance matrix $\bar{\MAT{K}}$ with analytic terms $\bar{K}_{ij}$ as derived in appendix \eqref{EQ:APPENDIX:APC:2DBALL} as follows:
\begin{equation}
\begin{split}
\bar{K}_{ij} =  \ABS{S(\nu, 0)}^2 \left \{  \begin{array}{cc}  \frac{2 J_1\PR{  D_{ij} \kappa \tau_r}} { D_{ij} \kappa \tau_r }, & D_{ij} \kappa \tau_r > 0 \vspace{2px} \\ 1, & D_{ij} \kappa \tau_r = 0 \end{array} \right .  ,
\end{split}
\label{EQ:DC:ALC:ACOUSTIC_COV}
\end{equation}
where $D_{ij} = \NORM{\VEC{v}_i - \VEC{v}_j}$ is the distance between loudspeaker unit-directions, and $J_1(z)$ is the Bessel function of the first kind. Note that at the listener location $\VEC{r} = \VEC{0}$, the normalized loudspeaker transfer functions are constant in \eqref{EQ:PT:CAUSAL_FILT}. Thus, the acoustic covariance matrix $\bar{\MAT{K}}$ degenerates to the rank-1 matrix $\mathring{\MAT{K}} = \ABS{S(\nu, 0)}^2 \VEC{1} \VEC{1}^T$ as the evaluation radius decreases to zero in $\lim_{\tau_r \rightarrow 0} g(\VEC{r}) = \delta$.  We therefore decompose the acoustic covariance as follows:


Let the acoustic covariance matrix in \eqref{EQ:DC:ALC:ACOUSTIC_POW} be a mixture of the listener location, and listening area covariances given by
\begin{equation}
\begin{split}
\MAT{K}  = (1 - \alpha)  \mathring{\MAT{K}} + \alpha \bar{\MAT{K}}, \quad
0 \leq \alpha \leq 1, 
\end{split}
\label{EQ:DC:ALC:ACOUSTIC_COV_GEN}
\end{equation}
where the acoustic covariance for $\alpha = 0$ evaluates only the direct acoustic transfer function from loudspeakers to the listener location. The quadratic constraints in \eqref{EQ:DC:ALC:ACOUSTIC_POW} linearize to $ \VEC{x}^T \VEC{1} = \sqrt{\rho} \ABS{S(\nu, 0)}^{\minus 1}$ for non-negative $\VEC{x}$;  maximizing $\lambda$ s.t. the linear gain summation constraint maximizes the panning sensitivity. Conversely, the acoustic covariance for $\alpha = 1$ evaluates the acoustic transfer functions over a larger listening area; maximizing $\lambda$ s.t. the quadratic equality constraint maximizes panning efficiency. Moreover, the loudspeaker acoustic covariances in the listening area at the limits are correlated in low-frequency $\lim_{\kappa \rightarrow 0} \bar{\MAT{K}} =  \mathring{\MAT{K}}$, and uncorrelated in high-frequency or large evaluation radii $\lim_{\kappa \rightarrow \infty} \bar{\MAT{K}} = \lim_{\tau_r \rightarrow \infty} \bar{\MAT{K}} = \MAT{I}$. Therefore, the mixture of covariances \eqref{EQ:DC:ALC:ACOUSTIC_COV_GEN} are  proportional to $\MAT{K} \propto  (1 - \alpha) \VEC{1} \VEC{1}^T + \alpha \MAT{I}$. We now formulate the loudspeaker steering optimization w.r.t. spatial, electrical, and acoustical constraints.

\textbf{Optimal Panning Sensitivity and Efficiency (OPSE):}
Maximizing the panning sensitivity $\lambda$ subject to spatial, acoustical, and electrical constraints is the second-order cone problem \citep{ALIZADEH_2003_SECOND_CONE} given by
\begin{equation}
\begin{split}
(\lambda_*, \VEC{x}_*) & = \ARGMAX{\lambda. \VEC{x}} \, \lambda \qquad \textrm{s.t.} \quad 
\lambda \geq 0, \\
\MAT{V} \VEC{x} & = \lambda \VEC{s}, \quad 
\VEC{x}^T \MAT{K} \VEC{x} \leq \rho, \quad 
\VEC{0} \leq \VEC{x} \leq \VEC{\tau}, 
\end{split}
\label{EQ:DC:OBJ}
\end{equation}
where a feasible solution always exist if the acoustic loudness's equality constraint in \eqref{EQ:DC:ALC:ACOUSTIC_POW} is relaxed to be in inequality; acoustic loudness is tight w.r.t. $\rho$ if panning sensitivity \eqref{EQ:DC:SDC:SENSITIVITY} or efficiency \eqref{EQ:DC:SDC:EFFICIENCY} is also maximized. We can eliminate $\lambda$ by left-multiplying both sides of the equality constraints in \eqref{EQ:DC:OBJ} by unit-direction $\VEC{s}^T$ to yield $\lambda = \VEC{s}^T \MAT{V} \VEC{x}$, and the equality constraint matrix $\MAT{A} =  (\MAT{I} - \VEC{s} \VEC{s}^T) \MAT{V}$. The equivalent optimization in only $\VEC{x}$ is expressed as follows:
\begin{equation}
\begin{split}
\VEC{x}_*  & = \ARGMAX{\VEC{x}} \, \VEC{c}^T \VEC{x} \qquad \textrm{s.t.} \quad 
\VEC{c}^T \VEC{x} \geq 0, \\
\MAT{A} \VEC{x} & = \VEC{0}, \quad 
\VEC{x}^T \MAT{K} \VEC{x} \leq \rho, \quad 
\VEC{0} \leq \VEC{x} \leq \VEC{\tau}, 
\end{split}
\label{EQ:DC:OBJ_EQUIV}
\end{equation}
where the objective maximizes the panning gains $\VEC{x}$ in the direction of vector $\VEC{c} =  \MAT{V}^T \VEC{s}$, consisting of cosine similarities between the target and loudspeaker unit-directions. Moreover, the equality constraints restrict $\VEC{x}$ to the null space of $\MAT{A}$, which has nullity $N-1$. Thus for real-time applications and small number of loudspeakers $(N \leq 5)$, we remove the equality constraints and reduce the number of variables via the linear transformation of the panning gains $\VEC{x} = \bar{\MAT{A}} \VEC{y}$ along an orthonormal basis $ \bar{\MAT{A}}^T \bar{\MAT{A}}  = \MAT{I}$ of the null space $\bar{\MAT{A}} \in  \textrm{span}\PR{ \textrm{ker} \PR{ \MAT{A}}} \in \FIELD{R}^{N \times N-1}$. The optimization in the kernel space reduces to linear and quadratic inequality constraints given by
\begin{equation}
\begin{split}
 \VEC{y}_* & = \ARGMAX{\VEC{y}} \, \bar{\VEC{c}}^T \VEC{y} \quad \textrm{s.t.} \,\,\, 
\begin{array}{c}  \bar{\VEC{c}}^T \VEC{y} \geq 0, \\   \VEC{0} \leq \bar{\MAT{A}}\VEC{y} \leq \VEC{\tau},  \end{array} \,\,\,
  \VEC{y}^T  \bar{\MAT{K}} \VEC{y} \leq \rho,
\end{split}
\label{EQ:DC:OBJ_EQUIV_KER}
\end{equation}
where $\bar{\VEC{c}} =  \bar{\MAT{A}}^T \VEC{c}$, and $\bar{\MAT{K}} = \bar{\MAT{A}}^T \MAT{K} \bar{\MAT{A}}$, and the feasible region is convex. Lastly, the steering direction $\VEC{s}$ can be infeasible where only the trivial solution $\VEC{x} = \VEC{0}$ satisfies the VBAPS equality constraint; dropping the VBAPS constraints $\MAT{A} \VEC{x} = \VEC{0}$ and $\VEC{c}^T \VEC{x} \geq 0$ in the primary form \eqref{EQ:DC:OBJ_EQUIV} relaxes the feasible space to be convex. Therefore, optimal solutions for both the null space \eqref{EQ:DC:OBJ_EQUIV_KER} and relaxed primary forms can be efficiently found via interior-point methods. Let us now investigate the solutions to \eqref{EQ:DC:OBJ}, \eqref{EQ:DC:OBJ_EQUIV}, \eqref{EQ:DC:OBJ_EQUIV_KER} under various acoustic power, covariance, and loudspeaker layouts in practical applications. 

\section{Experiments}
\label{SEC:EXP}

\begin{figure}[h]
\begin{center}
  {\includegraphics[width=0.95\columnwidth]{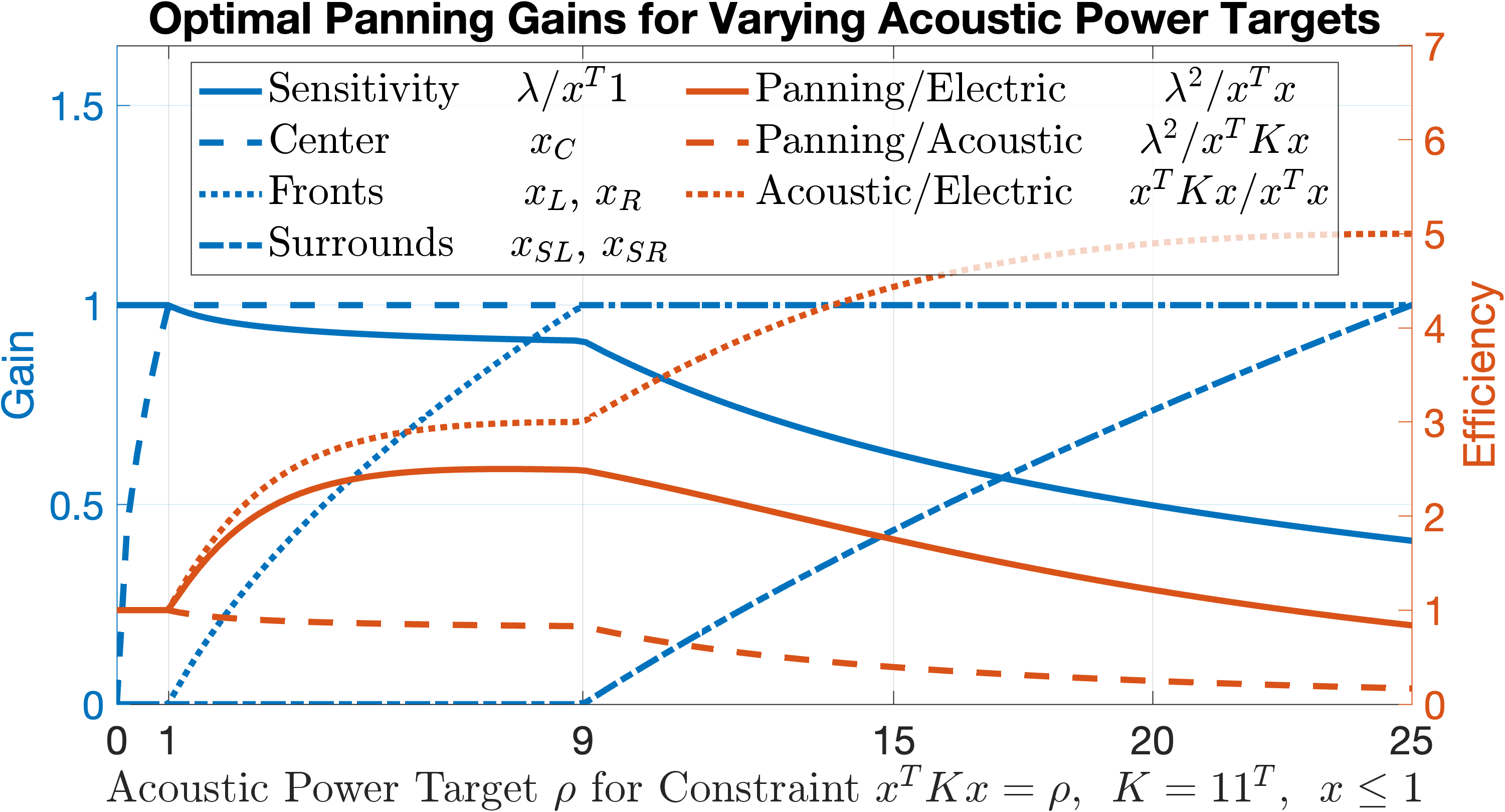}}
\caption{OPSE center content more uniformly distributes across $5.0$ ITU loudspeakers for increasing acoustic power targets $\rho$, and constant electrical headroom.}
\label{FIG:EXP:VARYACOSUTICPOW}
\end{center}
\end{figure}

\textbf{Distributed Center Channel:}
In the $5.0$ multichannel standard, the center content channel is fully sent to a center loudspeaker in a $5.0$ ITU layout (left = $-30^{\circ}$, right = $30^{\circ}$, center = $0^{\circ}$,  surround left = $-110^{\circ}$, surround right = $110^{\circ}$), where the maximum acoustic power (unity) is limited to that of a single loudspeaker. Under OPSE, we can specify a larger acoustic power target $\rho$ via the equality constraint $\VEC{x}^T \MAT{K} \VEC{x} = \rho$, spatial panning constraints of a center steering direction $\VEC{s} = \BK{1;  0}$, and unity electrical constraints $\VEC{x} \leq \VEC{1}$ WLOG. The optimal panning sensitivity gains for the listener location's acoustic covariance $\MAT{K} = \VEC{1} \VEC{1}^T$ are shown in  Fig. \ref{FIG:EXP:VARYACOSUTICPOW} for increasing acoustic power $\rho$ targets. For acoustic power targets $0 < \rho \leq 1$, only the center loudspeaker is active $0 < x_C \leq 1$, and  panning sensitivity is maximum. For $1 < \rho \leq 9$, the center loudspeaker exhausts its headroom and the left and right loudspeakers equally engage $(0 < x_{L, R} \leq 1, \, x_C = 1)$, resulting in a slight loss in panning sensitivity ($0.9$ at $\rho = 9$), and increase in both panning/electric and acoustic/electric efficiency. For $9 < \rho \leq 25$, the left and  right loudspeakers exhausts their headroom and the surround loudspeakers equally engage $(0 < x_{SL, SR} \leq 1, \, x_{L, R, C} = 1)$, resulting in a sharper loss to panning sensitivity and degradation to panning/electric efficiency as the center steering direction lies in the infeasible sector of the surround loudspeaker pair. Note that for inequality constraints $\VEC{x}^T \MAT{K} \VEC{x} \leq \rho$, the surround panning gains remain in-active as the quadratic constraint is not tight for $\rho > 9$. Panning sensitivity therefore monotonically decreases for larger acoustic power targets.

\begin{figure}[h]
\begin{center}
  {\includegraphics[width=0.95\columnwidth]{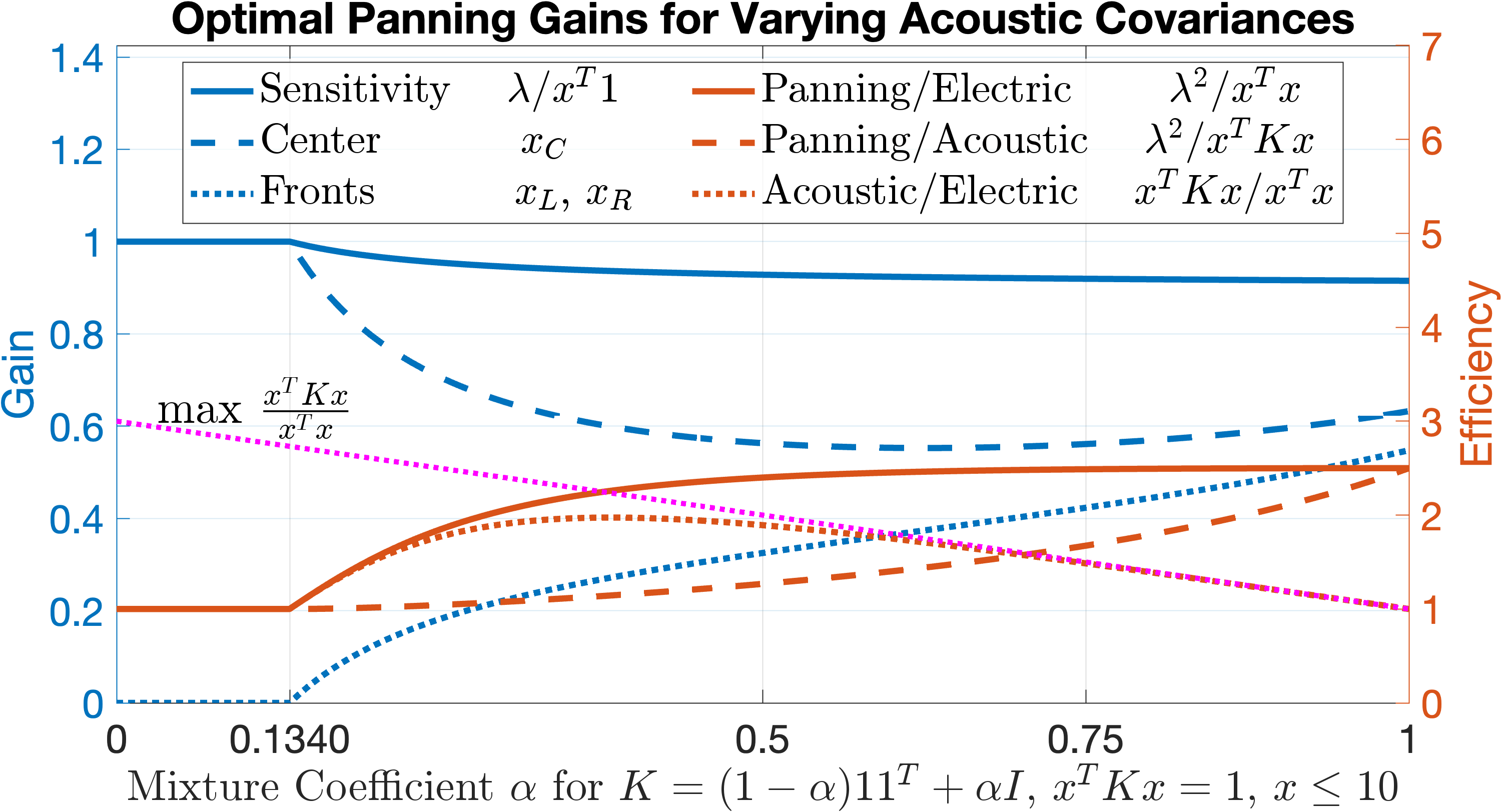}}
\caption{OPSE center content gains for $3.0$ ITU loudspeakers converge to the acoustic/electric Rayleigh quotient maximizer in diffuse-field conditions.}
\label{FIG:EXP:VARYALPHA}
\end{center}
\end{figure}

\textbf{Diffuse-field Panning:} In reverberant environments, acoustic covariance between well-separated loudspeakers in the listening area decreases due to increasing variations in acoustic reflection path responses. Normalized loudspeakers produce a mixture of correlated sound-fields from their direct acoustic paths, and less correlated diffuse-fields from their reflection paths over a listening area. The acoustic covariance in the listening area is therefore proportional to \eqref{EQ:DC:ALC:ACOUSTIC_COV_GEN}. Let us reconsider the previous case of distributed center channel over a $3.0$ ITU layout (left = $-30^{\circ}$, right = $30^{\circ}$, center = $0^{\circ}$). Under OPSE, we constrain the acoustic power to unity $\VEC{x}^T \MAT{K} \VEC{x} = 1$, relax the electrical headroom $\VEC{x} \leq \VEC{10}$, and vary the mixture of acoustic covariances as shown in Fig. \ref{FIG:EXP:VARYALPHA}. For correlated sound-fields  $0 \leq \alpha \leq 1 - \VEC{s}^T \VEC{v}_{L}$, only the center loudspeaker is active as panning sensitivity is maximum. For less correlated sound-fields $1 - \VEC{s}^T \VEC{v}_{L} < \alpha \leq 1$, the center loudspeaker attenuates relative to the left and right loudspeakers as more uniform-distributed gains yield both higher acoustic/panning and panning/electric efficiency. The gap between acoustic/electric efficiency and its theoretical Rayleigh quotient maximum, given by the largest eigenvalue of $\MAT{K}$, closes at the diffuse-field limit $\alpha = 1$. OPSE therefore converges to the largest eigenvector of $\MAT{K}$ under diffuse-field conditions where source-localization is difficult.

\begin{figure*}[h]
\begin{center}
  {\includegraphics[width=0.95\textwidth]{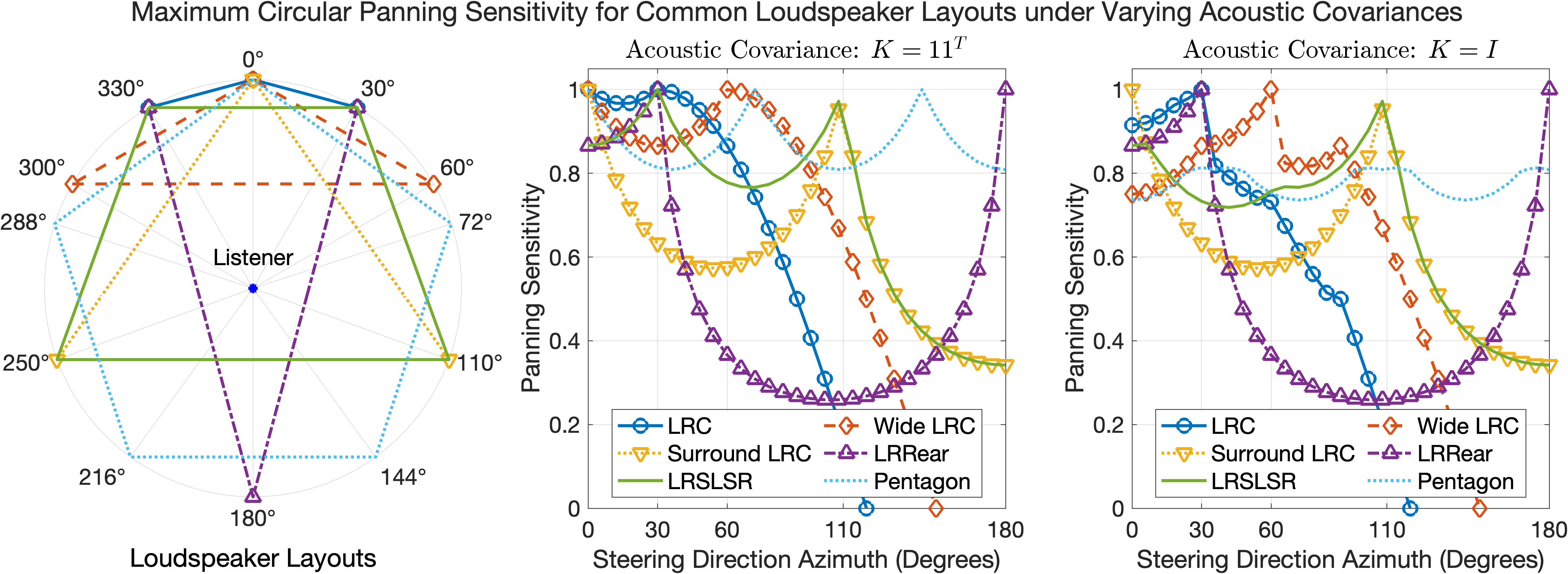}}
\caption{Maximum panning sensitivity in \eqref{EQ:DC:OBJ_EQUIV} varies across azimuth steering directions for different loudspeaker layouts, and in anechoic and diffuse-field conditions. All triangle arrangements have low-sensitivity gaps. A minimum of five uniform-spaced loudspeakers (Pentagon) achieves moderate sensitivity in diffuse-field conditions across all of azimuth.}
\label{FIG:EXP:LAYOUT}
\end{center}
\end{figure*}

\textbf{Circular Panning Across Loudspeaker Layouts:} For adaptive multichannel reproduction, it is desirable to render content channels over common loudspeaker layouts shown in Fig. \ref{FIG:EXP:LAYOUT} for any listener location and front-direction. Under OPSE, we can evaluate the panning sensitivity for all steering directions in azimuth in both anechoic $\MAT{K} = \VEC{1} \VEC{1}^T$ and diffuse-field $\MAT{K} = \MAT{I}$ conditions. Let us constrain the acoustic power to unity $\VEC{x}^T \MAT{K} \VEC{x} = 1$, relax the electrical headroom $\VEC{x} \leq \VEC{10}$, and vary $\VEC{s} = [\cos \theta; \, \sin \theta ]$ for the half-circle $0 \leq \theta \leq \pi$ as the layouts are symmetric w.r.t. $\theta = 0$. For layouts with only frontal loudspeakers such as LRC, and wide LRC, the panning sensitivity remains high  $ > 0.85$ for feasible steering directions. For infeasible steering directions, the VBAPS constraints are dropped in \eqref{EQ:DC:OBJ_EQUIV}, and the panning sensitivity, taken to be $\VEC{c}^T \VEC{x} / \VEC{x}^T \VEC{1}$, decrease for larger $\theta$. The solutions are continuous w.r.t. $\theta$ for the anechoic covariance but discontinuous for the diffuse-field covariance at the feasibility boundary of $\theta$. For triangular loudspeaker layouts (surround LRC, LRRear) containing the listener, only $2/3$ loudspeakers are active for any given $\theta$. The solutions therefore uniquely satisfy the VBAPS constraints and are equivalent in both anechoic and diffuse-field conditions. LRRear has acceptable panning sensitivity between $\ABS{\theta} \leq 30^{\circ}$, but minimal panning sensitivity near surround steering angles $100 \leq \theta \leq 110$. Surround LRC has low panning sensitivity for the left and right steering angles $\theta = \pm 30^{\circ}$. For the LRSLSR layout, the panning sensitivity degrades in diffuse-field  conditions for frontal angles $\ABS{\theta} \leq 60^{\circ}$, and is minimal in the surround loudspeaker pair's gap $110^{\circ} \leq \theta \leq  250^{\circ}$. For the pentagon layout of uniformly spaced loudspeakers, anechoic and diffuse-field conditions have acceptable  $> 0.8$ and borderline  $> 0.7$ panning sensitivity respectively, with the latter also having lower variance. Under OPSE, the pentagon layout is therefore suited for uniform directional circular panning, LRSLSR for non-rear directional panning, and wide LRC for frontal to semi-surround directional panning for content reproduction.

\section{Discussion}
\label{SEC:DISC}

While Bayesian loudspeaker normalization and OPSE formulations are acoustic measurement-free sound-field correction methods, their accuracy may be improved with indirect acoustic measurements. Loudspeaker acoustic covariances at the listening area, critical distances, and distance attenuation rates may be estimated from simplified room acoustic models such as image-sources \citep{lehmann2008prediction} if the room dimensions are known, and from loudspeaker-to-microphone acoustic transfer functions co-located on supporting smart-loudspeaker devices.  In such instances, we may couple loudspeaker normalization with OPSE by substituting the in-situ loudspeaker acoustic covariance estimates in-place of anechoic and planewave mixture. Extension to 3D layouts can be considered for some loudspeaker arrangements, but may be impractical for satisfying VBAPS constraints and ensuring feasible coverage over spherical coordinates.

\section{Conclusion}
\label{SEC:CONC}

We presented a loudspeaker filtering method that normalizes multiple loudspeakers to a common acoustic target for a non-stationary listening location or area. Loudspeaker normalization angles w.r.t. the listener location were adapted via Bayesian posteriors over circular distribution probability density functions. We then formulated panning gain optimization problems by relaxing VBAP constraints to give a novel panning sensitivity / efficiency objective, and specified electrical, acoustical domain constraints. Augmented, primary, null-space, and relaxed forms of the OPSE problem were derived. Lastly, practical experiments quantified the OPSE solutions in applications for distributing excess content channels, evaluating efficiency in anechoic to diffuse-field conditions, and recommending loudspeaker layouts for circular panning and multichannel content reproduction.

\section{Appendix}

\textbf{Circular Distribution:}
We can evaluate $P(-a \leq \theta \leq a)$ for $f(\theta)$ in \eqref{EQ:PT:PDF} via the power-series approximation and cosine-integral \citep{WEISSTEIN_COS_INT} given by
\begin{equation}
\begin{split}
\int_{\minus a}^{a} e^{\frac{\minus d^2(\theta)}{2\ell^2} } d \theta 
& = \int_{\minus a}^{a} e^{\frac{\cos (\theta) \minus 1}{\ell^2}}  d \theta 
 =  \sum_{n=0}^\infty  \frac{\int_{\minus a}^{a} \cos^n (\theta) d \theta   \minus 1 }{\ell^{2n}  n!}.
\end{split}
\label{EQ:APPENDIX:CD:INT}
\raisetag{9.5ex}
\end{equation}
The product of circular distributions is a circular distribution as the sum of weighted and phase-shifted cosines in the exponents' terms is a cosine given by
\begin{equation}
\begin{split}
f_i(\theta) & \propto e^{\frac{\minus d^2(\theta - \mu_i)}{2\ell_i^2} } \propto
 e^{\frac{\cos(\theta - \mu_i)}{\ell_i^2 }  } \quad \Rightarrow \\
f_i(\theta) f_j(\theta) & \propto e^{\frac{\ell_j^2 \cos(\theta - \mu_i)  + \ell_i^2 \cos(\theta - \mu_j) }{\ell_i^2 \ell_j^2} } = 
 e^{ \frac{\cos (\theta - \mu )}{\ell^2} }, 
\end{split}
\label{EQ:APPENDIX:CD:PROD_CD}
\end{equation}
where the mean and dispersion of the product are given by
\begin{equation}
\begin{split}
\mu_{ij} & = \atantwo \PR{ \frac{ \sin (\mu_i)}{\ell_i^2 } + \frac{\sin(\mu_j)}{\ell_j^2  } , \,
\frac{\cos (\mu_i)}{ \ell_i^2  } + \frac{\cos(\mu_j)}{ \ell_j^2 }   }, \\
\ell_{ij}^2 & = \ell_i^2 \ell_j^2 \PR{\ell_i^4 + 2 \cos(\mu_i - \mu_j) \ell_i^2 \ell_j^2 + \ell_j^4 }^{\minus \frac{1}{2}}.
\end{split}
\label{EQ:APPENDIX:CD:PROD_CD_PRM}
\raisetag{3.5ex}
\end{equation}
The posterior dispersion depends on both prior mean and dispersion unlike that of a normal distribution. Substituting the empirical mean $\mu_i = \VEC{\bar{\theta}}^{\CR{t}}_n$, dispersion $\ell_i = \bar{\ell}_n^{\CR{t}}$, and the prior's mean  $\mu_j = \mu_n^{\CR{t - 1}}$, dispersion $\ell_j = \ell_n^{\CR{t-1}}$ in  \eqref{EQ:PT:PDF_STAT} gives the posterior mean $\mu_{ij} = \mu_n^{\CR{t}}$, dispersion $\ell_{ij} = \ell_n^{\CR{t}}$.



\textbf{Plane-wave Covariance:}
Let $p(\VEC{r}) = e^{\minus j \kappa \VEC{v}^T \VEC{r}}$ be the $2$D plane-wave equation with incident direction $\VEC{v}$, and the region of integration be the $2$D disc of radius $R$ where $\VEC{r} = \BK{x, y}^T$, $-R \leq x \leq R$, $y =  \sqrt{R^2 - x^2}$. We may express the incident angle and evaluation point in polar coordinates as follows:
\begin{equation}
\begin{split}
\VEC{v} & = \NORM{\VEC{v}} \BK{\cos \theta_v, \, \sin \theta_v}^T, \quad \VEC{r} = r \BK{\cos \theta, \, \sin \theta}^T,\\
\VEC{v}^T \VEC{r} & = \NORM{\VEC{v}} r \PR{\cos \theta_v \cos \theta + \sin \theta_v \sin \theta}  \\
& = \NORM{\VEC{v}} r \cos  \PR{\theta - \theta_v},  \\
-\VEC{v}^T \VEC{r} &= \NORM{\VEC{v}} r \cos  \PR{\pi - \theta + \theta_v}, \quad \textrm{ Cosine reflection}
\end{split}
\label{EQ:APPENDIX:APC:POLAR}
\end{equation}
where $0 \leq r \leq R$ and $0 \leq \theta \leq 2\pi$, which removes the dependence on the plane-wave incident angle in subsequent integrals. The first moment is analytic w.r.t. the Bessel function of the first kind $J_n(x)$ and given by
\begin{equation}
\begin{split}
\EXPECTATION{}{p(\VEC{r})} & = \frac{1}{\pi R^2}  \int_{0}^{2\pi} \int_{0}^{R}  r e^{-j k \VEC{v}^T \VEC{r} }  \, dr \, d\theta \\
& = \frac{1}{\pi R^2} \int_{0}^{R} r \int_{0}^{2\pi} e^{j k  \NORM{\VEC{v}} r \cos  \PR{\pi - \theta + \theta_v} } \, d\theta \, dr   \\
& = \frac{2}{R^2} \int_{0}^{R} r J_{0}( k \NORM{\VEC{v}}   r ) \, dr \quad \textrm{Hansen-Bessel \citep{TEMME_1996}} \\
& =  \frac{2 J_{1} (k \NORM{\VEC{v}}  R)}{k \NORM{\VEC{v}} R }. \quad \textrm{Bessel integral identity \citep{WEISSTEIN_BESSEL_FIRST_KIND}} \\
\end{split}
\label{EQ:APPENDIX:APC:2DBALL_MEAN}
\raisetag{15ex}
\end{equation}
Let $p_n(\VEC{r}) = e^{\minus j \kappa \VEC{v}_n^T \VEC{r}}$ be the plane-wave equation of the $n^{th}$ loudspeaker. The second moment can be expressed as $\VEC{v} = \VEC{v}_m - \VEC{v}_n$, which after substitution in \eqref{EQ:APPENDIX:APC:2DBALL_MEAN} follows
\begin{equation}
\begin{split}
\EXPECTATION{}{p_m(\VEC{r}) p_n^*(\VEC{r}) }   = \frac{1}{\pi R^2} \int_{0}^{2\pi} \int_{0}^{R}  r e^{\minus j k \PR{\VEC{v}_m \minus \VEC{v}_n}^T \VEC{r} }  \, dr \, d\theta \\
 = \left \{  \begin{array}{cc}  \frac{2 J_1\PR{k \NORM{\VEC{v}_m - \VEC{v}_n} R}} {k \NORM{\VEC{v}_m - \VEC{v}_n} R }, & \NORM{\VEC{v}_m - \VEC{v}_n} > 0 \vspace{2px} \\ 1, & \NORM{\VEC{v}_m - \VEC{v}_n} = 0 \end{array} \right . .
\end{split}
\label{EQ:APPENDIX:APC:2DBALL}
\end{equation}

\bibliographystyle{jaes}

\bibliography{IEEEabrv, refs}

\end{document}